\documentclass[dvips]{article}
\usepackage{icrctc07}
\begin{document}

\title{Stellar Photon and Blazar Archaeology with Gamma-rays}
\shorttitle{Photon and Blazar Archaeology}
\authors{Floyd W. Stecker$^{1}$}
\shortauthors{Floyd W. Stecker}
\afiliations{$^1$NASA  Goddard  Space  Flight  Center,  Greenbelt,  MD
20771, USA} \email{stecker@milkyway.gsfc.nasa.gov}

\abstract{Ongoing  deep  surveys  of  galaxy  luminosity  distribution
functions,  spectral  energy  distributions  and  backwards  evolution
models  of star  formation rates  can be  used to  calculate  the past
history of intergalactic photon  densities and, from them, the present
and  past optical  depth of  the universe  to $\gamma$-rays  from pair
production interactions with these photons.  Stecker, Malkan \& Scully
have  recently calculated  the densities  of  intergalactic background
light (IBL)  photons of energies  from 0.03 eV  to the Lyman  limit at
13.6  eV and  for redshifts  0$ <  z <$  6, using  deep  survey galaxy
observations  from the  {\it Spitzer,  Hubble} and  {\it  GALEX} space
telescopes.   From  these  results,  they  have  predicted  absorption
features for blazar spectra.

This proceedure can  also be reversed by looking  for sharp cutoffs in
the spectra of extragalactic $\gamma$-ray sources at high redshifts in
the multi-GeV energy range with  {\it GLAST} (the Gamma-ray Large Area
Space  Telescope).  Determining  the cutoff  energies of  sources with
known  redshifts and  little intrinsic  absorption may  enable  a more
precise determination of  the IBL photon densities in  the past, i.e.,
the "archaeo-IBL",  and therefore  allow a better  measure of  the past
history of the total star formation rate, including that from galaxies
too faint to be observed.

Conversely,  observations   of  sharp  high  energy   cutoffs  in  the
$\gamma$-ray  spectra of  blazars  at unknown  redshifts  can be  used
instead of spectral lines to  give a measure of their redshifts. Also,
given  a knowledge  of the  archaeo-IBL, one  can derive  the intrinsic
$\gamma$-ray  spectra and  luminosities  of blazars  over  a range  of
redshifts and  look for possible trends in  blazar evolution. Stecker,
Baring \& Summerlin have found  some evidence hinting that TeV blazars
with   flatter  spectra   have  higher   intrinsic   TeV  $\gamma$-ray
luminosities  and  indicating  that  there  may be  a  correlation  of
flatness and  luminosity with redshift.  {\it GLAST}  will observe and
investigate many  blazars in the  GeV energy range and  will therefore
provide much new information regarding this possibility.}  \maketitle




\section{Introduction}

Space-based telescopes  that are  dedicated to exploring  the Universe
out to large  distances in wavelength ranges from  the far-infrared to
the X-ray  range are now in  place and the Gamma-ray  Large Area Space
Telescope, {\it  GLAST}, is  scheduled to be  early next  year.  These
facilities  are capable  of probing  the Universe  to study  the early
primordial phases of star  formation and galaxy evolution.  Studies of
cosmic photons  from NASA space observatories  supply important inputs
as well into astrophysics, physics and cosmology.  Presently, the {\it
Spitzer} space infrared telescope  facility is probing deeply into the
past to  study galaxy  formation and evolution  in the  infrared. {\it
GALEX} is making similar  observational studies in the ultraviolet. In
the  $\gamma$-ray  energy range  {\it  SWIFT}  is gathering  important
information on $\gamma$-ray  bursts back to the distant  past and {\it
GLAST}  will study  sources  and diffuse  fluxes  of $\gamma$-rays  at
energies up to  100 GeV.  The astronomical era  of the ``deep fields''
has arrived.

Whereas archaeological  sites such as  Chi-chen Itza, which many  of us
visited  on  the  conference  excursion,  allow us  to  peer  back  to
$\sim$kyr in the past, the deep astronomical surveys from the radio to
the $\gamma$-ray wavelengths  allow us to peer back  $\sim$Gyr or more
into the history of the Universe.

\section{Calculating Intergalactic Archaeophoton Densities}

Stecker, Malkan \& Scully ~\cite{sms} have used the approach pioneered
in Refs.~\cite{ms98} and  ~\cite{ms01} to calculate past intergalactic
infrared photon  fluxes and densities  as they evolved along  with the
galaxies  that produced  them.  This  method, a  "backwards evolution"
scheme, is  an empirically based  calculation which uses as  input (1)
the luminosity  dependent galaxy spectral  energy distributions (SEDs)
based   on  observations   of   normal  galaxy   infrared  SEDs,   (2)
observationally based  galaxy luminosity distribution  functions (LFs)
and  (3)  redshift dependent  galaxy  luminosity evolution  functions.
These  are  empirically  derived  curves  giving  the  universal  star
formation rate or luminosity density  as a function of redshift, which
are referred  to as Lilly-Madau plots.  Ref.~\cite{sms}  uses new deep
survey observations,  primarily from the {\it  Spitzer} infrared space
observatory,  to make  improvements  to the  previous calculations  of
Refs.~\cite{ms98} and ~\cite{ms01}.

\begin{figure}
  \includegraphics[height=.33\textheight]{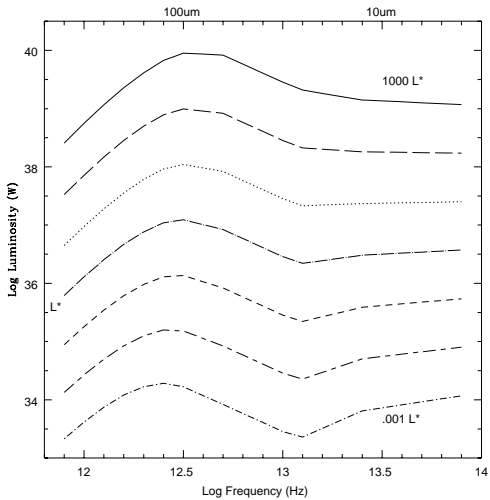}
\caption{Empirically derived galaxy SEDs for various luminosities from
$10^{-3}L_{*}$ to $10{^3}L_{*}$.}
\label{galspec}
\end{figure}

\subsection{Galaxy Infrared SEDs as a Function of Luminosity}

The key  empirically supported  assumption made in  Ref.~\cite{sms} is
that the luminosity of a  galaxy determines the average galaxy SED and
that therefore the galaxy luminosity distribution function (LF) can be
predicted statistically  from its observed luminosity  in one infrared
waveband,   here  chosen   to  be   60$\mu$m~\cite{sa90}.   It   is  a
well-established  fact that  more luminous  galaxies (now  and  in the
past) have  higher rates of ongoing  star formation and  that the star
formation rate was higher in the past.

Empirically, it  is also  found that for  the more  luminous galaxies,
relatively more  of the energy from  these young stars  is absorbed by
dust  grains and re-radiated  in the  thermal infrared;  more luminous
galaxies have higher infrared flux relative to optical flux and warmer
infrared spectra.   These clear luminosity-dependent  trends in galaxy
SEDs were well  determined locally from the combination  of {\it IRAS}
(Infrared Astronomy  Satellite) and ground-based  photometry for large
({\it  e.g.}, all sky)  samples. The  infrared SEDs  as a  function of
galaxy  luminosity   used  in  \cite{sms}  were   based  on  broadband
photometry of  {\it IRAS}  selected samples and  show the  average SED
emission trends discussed above. The  family of average galaxy SEDs at
various   luminosities    in   terms   of   $L_{*}$    is   shown   in
Figure~\ref{galspec}, taken from  Ref.~\cite{ms98}, where $L_{*} = 8.4
\times 10^{23}$ W Hz$^{-1}$ is the luminosity at 60 $\mu$m.

Other computations of infrared backgrounds and source counts have used
different SEDs,  based on somehwat different combinations  of data and
models, in some cases estimated  in more spectral detail.  One may ask
whether   these   new  SEDs   might   differ   from   those  used   in
Refs.~\cite{ms98},~\cite{ms01}   and~\cite{sms},  either   in  overall
colors,  or  in  detail  around  the  7--12$\mu$m  region,  where  the
strongest spectral features from polycyclic aromatic hydrocarbon (PAH)
emission and  silicate emission are  found. The best example  of these
new galaxy template  SEDs can be found in  Ref~\cite{xu}.  As shown in
Ref.~\cite{sms},   the  agreement   between  the   infrared   SEDs  in
Refs.~\cite{xu}  and  ~\cite{sms}  is  excellent.   As  long  as  this
agreement  holds  for  SEDs  of  galaxies near  the  ``knee''  of  the
approximate broken power-law LFs  is reasonable (see Figure \ref{LF}),
the  final computed  intergalactic photon  densities will  also agree,
because they are dominated  by galaxies with luminosities $\sim L_{*}$
around the knee.

\subsection{The Local Infrared Luminosity Function}

The foundation of the backwards evolution calculation~\cite{sms} is an
accurately   determined   local   infrared  luminosity   function   of
galaxies. We  used the LF observationally  derived in Ref.~\cite{sa90}
because  it was  based on  an extensive  analysis of  very  large data
sample  of  galaxies. This  LF  was updated  by  using  the even  more
thorough local infrared LF  given in the more recent Ref.~\cite{ta03}.
A   low-luminosity   power-law  index   of   -1.35   was  adopted   in
Ref.~\cite{sms}  for  the   broken  power-law  differential  LF  which
steepened by 2.25 at high  luminosities.  This LF takes better account
of   the   large   number   of   fainter   galaxies   that   are   now
known~\cite{bl05}. The cosmological values adopted in Ref. ~\cite{sms}
were $h =  0.7$ and a $\Lambda$CDM cosmology  with $\Omega_{\Lambda} =
0.7$ and $\Omega_{m} = 0.3$.

\begin{figure}
  \includegraphics[height=.33\textheight]{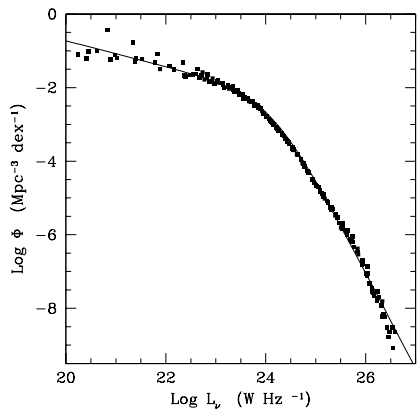}
\caption{60$\mu$m local galaxy luminosity function.}
\label{LF}
\end{figure}

Figure \ref{LF} compares the broken power-law  LF at $ z = 0$ with the
data given in Ref.~\cite{ta03}.  Note that the luminosity $L_{*}$ used
in Figure ~\ref{galspec}  is defined to be the  luminosity at the knee
of the LF shown in Figure ~\ref{LF}. It can be seen that the agreement
between the analytic curve and the observational data is excellent.

\subsection{Evolution of the IBL SED with Redshift}

It is now well known that  galaxies had a brighter past owing to their
higher rates of  star formation and the fading  of stellar populations
as  they  age.   The   simplest  resulting  evolution  of  the  galaxy
luminosity function  is a  uniform shift in  either the  vertical axis
(number  density evolution),  or  in the  horizontal axis  (luminosity
evolution.)   For a  pure  power-law luminosity  function, number  and
luminosity  evolution  are  mathematically  equivalent.   In  reality,
however,  to  avoid unphysical  divergences  in  the  total number  or
luminosity of galaxies, the luminosity function must steepen at high L
and  flatten  at  low L.   Thus  real  LF's  will  have at  least  one
characteristic ``knee"  separating the  steep high-L portion  from the
flatter low-L  slope.  For  typical LF's this  results in most  of the
luminosity being emitted  by galaxies within an order  of magnitude of
this knee.   Thus large  uncertainties and errors  in the LF  far from
this knee will  hardly change most of the  results ({\it e.g.}, number
counts and integrated diffuse backgrounds).

Strong  luminosity evolution  of galaxies,  {\it i.e.},  a substantial
increase in the luminosity of this knee with redshift, is consistently
found by  many observations relating  infrared luminosity to  the much
higher  star  formation  rate  at   $z  \sim  1$  and  to  the  recent
determination that  most ultraviolet-selected  galaxies at $z  \sim 1$
are also luminous infrared galaxies.

The exact  form of  the luminosity evolution  of galaxies is  the most
uncertain  input to  the  calculation  of the  IBL  density, since  it
depends on high redshift data.  Therefore, two plausible cases of pure
luminosity  evolution  were adopted  in  Ref.~\cite{sms}  in order  to
bracket  the   uncertainties  from  the  deep  surveys:   (1)  a  more
conservative  ``baseline'' (B)  evolution  scenario and  (2) a  ``fast
evolution'' (FE) scenario. A  more detailed discussion of these models
is given  in Ref.~\cite{sms}. The FE  model is favored  by recent {\it
Spitzer}  observations  \cite{lf,  pg}.   It also  provides  a  better
description  of  the  deep  {\it  Spitzer} number  counts  at  70  and
160$\mu$m than  the B model.   However, {\it GALEX}  (Galaxy Evolution
Explorer)  observations  indicate   that  the  redshift  evolution  of
ultraviolet  radiation  may  be  more  consistent  with  the  B  model
\cite{sch}. The {\it Spitzer  IRAC} (Infrared Array Camera) counts can
be best fit with an evolution  rate between these two models.  One way
of understanding  the somewhat smaller redshift evolution  of the star
formation  rate implied  by the  {\it GALEX}  ultraviolet observations
{\it vs.}  that obtained  from the {\it Spitzer} infrared observations
is that the effect of dust extinction followed by infrared reradiation
increases with redshift \cite{bu}.

\section{Calculation of the Intergalactic Background Light}

\begin{figure}
  \includegraphics[height=.33\textheight]{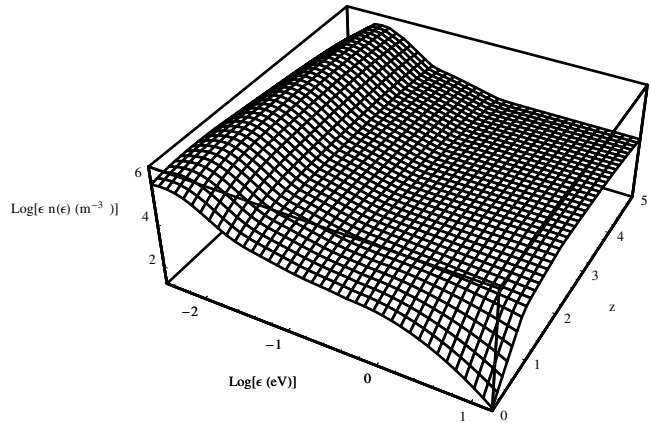}
\caption{The  photon   density  $\epsilon  n(\epsilon)$   shown  as  a
continuous function of energy and redshift.}
\label{3Dphot}
\end{figure}

\begin{figure}
  \includegraphics[height=.33\textheight]{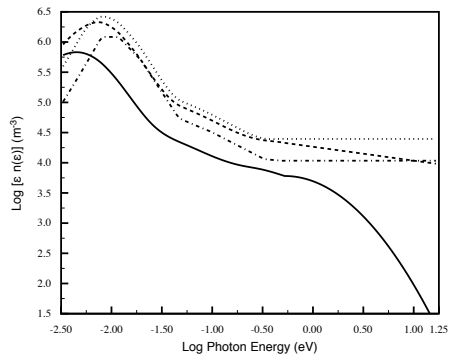}
\caption{The photon  density $\epsilon n(\epsilon)$ 
function of energy  for various redshifts based on  the fast evolution
model for  infrared evolution.  The  solid line is  for $ z =  0$, the
dashed line is  for $ z =  1$ , the dotted line  is for $ z  = 3$, the
dot-dashed line is for $ z = 5$.}
\label{photz}
\end{figure}

Our calculation of the  diffuse infrared background as described above
extends  up   to  a  rest   frequency  of  $\log  \nu_{Hz}   =  14.1.$
corresponding to an  energy of $\sim 0.5$ eV. This  is the location of
the peak in the spectral  energy distributions of most galaxies and is
produced by  the light  of red giant  stars.  The spectrum  (in energy
density  units) then  curves downward  rapidly to  higher frequencies,
with  modest dependence  on galaxy  luminosity.  Although  galaxy SEDs
have a peak at this energy,  this peak, as opposed to the far infrared
peak in galaxy  SEDs, only manifests itself as  an inflection point in
the  photon   density  spectrum.\footnote{We  note   that  the  energy
dependence    of   the    differential    photon   energy    spectrum,
$dn_{\gamma}/d\epsilon$, is obtained by dividing the SED by the square
of the photon energy ($\epsilon^2$)  so that the starlight ``peak'' in
the SED has very few photons  compared to the dust reradiation peak in
the   far  infrared.}    At   wavelengths  shortward   of  this   near
infrared-optical ``peak' the photon  density spectra drop steeply (See
Figures \ref{3Dphot} and \ref{photz} which plot $\epsilon n(\epsilon)$
for the case of the  fast evolution (FE) model.) The relative increase
in the ratio  of red-to-blue photons at the  higher redshifts shown in
these figures is produced by early generation, low luminosity stars.

\begin{figure}
  \includegraphics[height=.33\textheight]{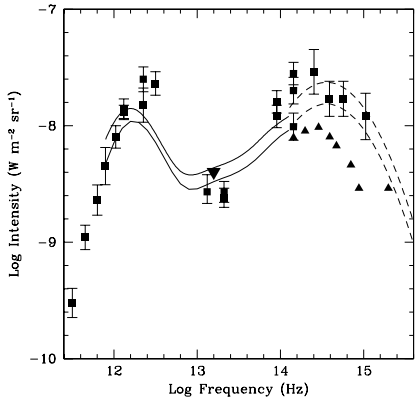}
\caption{Spectral  energy  distribution   of  the  diffuse  background
radiation  at $ z  = 0$,  sometimes referred  to as  the extragalactic
background light  (EBL). Error bars  show data points,  triangles show
lower limits  from number counts,  and the inverted triangle  shows an
upper limit  from TeV observations  ~\cite{sd97}. The upper  and lower
solid   lines  show   the  fast   evolution  and   baseline  evolution
predictions,  and the  dotted  lines show  their  extensions into  the
optical--ultraviolet, as described in Ref.~\protect\cite{ss98}.}
\label{IBL}
\end{figure}

 The low  redshift extragalactic background light  (EBL) SEDs obtained
are shown in Figure ~\ref{IBL}.   They are given for the two evolution
models previously  discussed, {\it  viz.}, the ``baseline''  (B) model
and the ``fast evolution'' (FE) model.

\section{Gamma-ray Absorption from Pair Production Interactions}

The potential importance of the photon-photon pair production process,
$\gamma \gamma  \rightarrow e^+ e^-$, in high  energy astrophysics has
been realized  for over 40 years  \cite{ni}.  It was  pointed out that
owing to interactions with the 2.7 K CMB, the universe would be opaque
to $\gamma$-rays  of energy above  100 TeV at  extragalactic distances
\cite{gs, j}.  If one  considers cosmological and redshift effects, it
was  further  shown that  photons  from  a  $\gamma$-ray source  at  a
redshift $z_{s}$  would be  significantly absorbed by  pair production
interactions with the CMB above an energy $\sim 100(1+z_{s})^{-2}$ TeV
\cite{st69, fs}.

Following  the discovery  by  the  {\it EGRET}  team  of the  strongly
flaring  $\gamma$-ray  blazar   3C279  at  redshift  0.54  \cite{h92},
Stecker, de Jager and Salamon \cite{sds} proposed that one can use the
predicted pair production absorption  features in blazars to determine
the intensity  of the infrared portion  of the IBL,  provided that the
intrinsic spectra  of blazars  extends to TeV  energies. It  was later
shown that the  IBL produced by stars in galaxies  at redshifts out to
$\sim 2$ would make the universe  opaque to photons above an energy of
$\sim 30$  GeV emitted  by sources  at a redshift  of $\sim  2$, again
owing to pair production interactions \cite{mp, ss98}.

As discussed  above, in Ref.~\cite{sms} this approach  was expanded by
using  recent data  from  the {\it  Spitzer}  infrared observatory  In
addition, data from  the {\it Hubble} deep survey  and the {\it GALEX}
mission were also used to determine the photon density of the IBL from
0.03 eV  to the Lyman  limit at  13.6 eV for  redshifts out to  6 (the
``archaeo-IBL'').\footnote{See  also Refs.   \cite{tt,  k1, k2}.}   The
results,  giving the  IBL photon  density  as a  function of  redshift
together with the  opacity of the CMB as a  function of redshift, were
then used  to calculate the  opacity of the universe  to $\gamma$-rays
for energies from 4 GeV to 100  TeV and for redshifts from $\sim 0$ to
5.

\begin{figure}
  \includegraphics[height=.33\textheight]{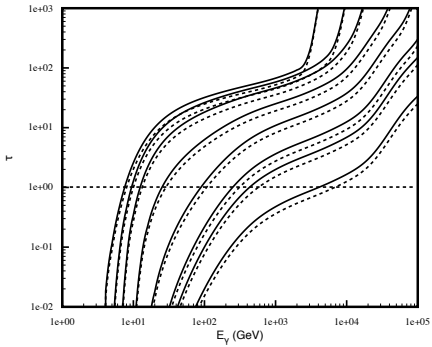}
  \caption{Optical  depth  of   the  universe  to  $\gamma$-rays  from
interactions  with photons  from  the  EBL and  2.7K  CMB for  various
redshifts, $z$.  The  solid lines are for the FE  model and the dashed
lines are  for the  B model.  The  curves shown  are for (from  top to
bottom)   $z$    =   5.0,   3.0,   2.0,   1.0,    0.5,   0.2,   0.117,
0.03\protect\cite{sms}.}
\label{tau}
\end{figure}

\begin{figure}
  \includegraphics[height=.33\textheight]{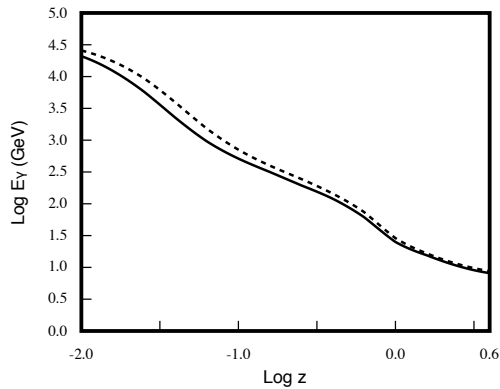}
\caption{A  Fazio-Stecker plot  (named  after the  paper  by Fazio  \&
Stecker  \protect\cite{fs})  which gives  the  critical optical  depth
$\tau = 1$  as a function of $\gamma$-ray energy  and redshift for the
fast evolution  (solid curve) and  baseline (dashed curve)  IGL cases.
Areas to  the right  and above these  curves correspond to  the region
where  the  universe  is  optically  thick  to  $  \gamma$-rays  (from
Ref.~\protect\cite{sms}.) }
\label{fsplot}
\end{figure}

The results of Ref. \cite{sms} as shown in Figure~\ref{tau} imply that
the universe  will become opaque  to $\gamma$-rays for sources  at the
higher redshifts  at somewhat  lower $\gamma$-ray energies  than those
given   in   Ref.   \cite{ss98}.    This   dependence   is  shown   in
Figure~\ref{fsplot}.  The  increased opacity calculated  at the higher
redshifts is because  the newer deep surveys have  shown that there is
significant  star formation  out to  redshifts $z  \ge  6$ \cite{bu04,
bo}),  greater  than  the value  of  $z_{max}  =  4$ assumed  in  Ref.
\cite{ss98}.  This conclusion is  also supported by recent {\it Swift}
observations of the redshift distribution of GRBs \cite{tj}.

\begin{table*}[t]
\begin{center}
{\bf Table  1.  Optical  Depth Parameters for  the Fast  Evolution and
Baseline Models}
\begin{tabular}{ccccc}
& & & & \\ \hline Evolution Model & A & B & C & D \\ \hline\hline Fast
Evolution & -0.475 & 21.6 & -0.0972 & 10.6 \\ Baseline & -0.346 & 16.3
& -0.0675 & 7.99 \\ \hline
\end{tabular}
\end{center}
\end{table*}

Stecker, Malkan  and Scully \cite{sms}  found that the  function $\tau
(E_{\gamma},  z)$  shown  in   Figure  ~\ref{tau}  can  be  very  well
approximated by the analytic form

\begin{equation}
\log \tau = Ax^4+Bx^3+Cx^2+Dx+E
\end{equation}

\noindent over  the range  $0.01 <  \tau < 100$  where $x  \equiv \log
E_{\gamma}$ (eV).  The correct coefficients  A through E are given for
various redshifts in  Ref. \cite{cor}. (This is the  corrected form of
the original table given in Ref. \cite{sms}.)

\section{Photon Archaeology}
  
\begin{figure}[h]
\begin{center}
  \includegraphics[height=.15\textheight]{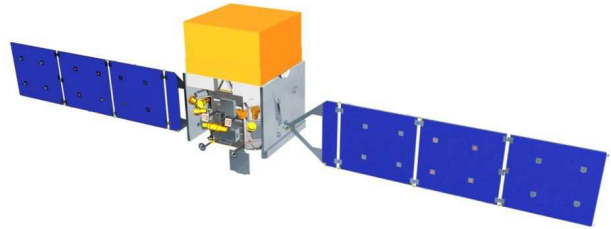}
\end{center}
\hspace{1.cm}
\caption{Schematic of the {\it GLAST} satellite deployed in orbit. The
{\it LAT}, which  comprises of a cluster of  16 silicon strip trackers
and a  calorimeter consisting of 8  layers of CsI(Tl)  crystals, is in
the  top (yellow)  area; the  Gamma Ray  Bust Monitors  {\it  GBM} are
located directly below.}
\label{glast}
\end{figure}

Observations of relatively nearby blazars using the {\it H.E.S.S.} and
{\it MAGIC} air \v{C}erenkov  telescopes have produced many results in
the TeV energy range. However,  as can be seen from Figure ~\ref{tau},
the critical energy range  for exploring sharp absorption cutoffs from
distant blazars is the multi-GeV range.  This is an energy range which
the  upcoming  {\it GLAST}  mission  will  explore.   The {\it  GLAST}
satellite (see  Figure ~\ref{glast}) will have a  large area telescope
called  the  {\it LAT}  which  is  designed  to study  cosmic  diffuse
$\gamma$-rays  and $\gamma$-ray  sources in  the energy  range between
$\sim$20 MeV and $> 300$  GeV (For more information about {\it GLAST},
go  to   {\tt  http://glast.stanford.edu/}.   By   using  {\it  GLAST}
observations to  determine the  absorption cutoff energies  of sources
with known redshifts caused by interactions of GeV range $\gamma$-rays
with low energy photons of the IBL, we can refine our knowledge of the
IBL photon  densities in the  past, {\it i.e.}, the  {\it archaeo-IBL},
and therefore  get a better  measure of the  past history of  the {\it
total} star  formation rate.   Conversely, observations of  sharp high
energy  cutoffs in  the  $\gamma$-ray spectra  of  sources at  unknown
redshifts can be  used instead of spectral lines to  give a measure of
their redshifts.   {\it GLAST}  must also investigate  the possibility
that intrinsic  absorption within some high  luminosity, high redshift
$\gamma$-ray sources may mimic absorption by the IBL ~\cite{re07}.  If
this is the case, such an effect will need to be disentangled from IBL
absorption.  One  key question is  how far the region  of $\gamma$-ray
emission is from the radiation field surrounding the black hole. There
evidence in the  case of M87 that the  $\gamma$-ray emission region is
greater than  120 pc from  the black hole~\cite{ch07}  arguing against
intrinsic absorption. However, this is only one source and it is not a
typical  blazar.   Observationally,  intrinsic  absorption  in  active
galactic  nuclei can be  investigated by  determining the  lower limit
envelope on  the total  absorption from  a set of  sources in  a known
narrow redshift range. This can be accomplished by {\it GLAST}.

\begin{figure}
  \includegraphics[height=.33\textheight]{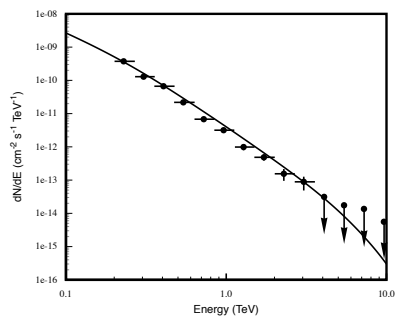}
\caption{The $\gamma$-ray data from the {\it H.E.S.S.} for PKS2155-304
compared  with  the theoretically  obtained  spectrum for  PKS2155-304
calculated by  assuming an  intrinsic source spectrum  proportional to
$E^{-2}$ and multiplying by  $e^{-\tau}$ using $\tau(z=0.117)$ for the
FE model of Ref.~\protect\cite{sms}. }
\label{PKS2155}
\end{figure}

\section{Blazar Spectra and Luminosities, Now and Then}

As an example of the application of the $\gamma$-ray absorption models
derived  in Ref.~\cite{sms}, Figure  \ref{PKS2155} shows  the absorbed
spectrum of the  blazar PKS 2155-304 at $z =  0.117$ assuming that its
intrinsic  spectrum  in  the  energy  range shown  is  an  approximate
$E^{-2}$  power  law. This  spectrum  is  then  compared it  with  the
spectrum  observed by  the {\it  H.E.S.S.} TeV  $\gamma$-ray telescope
array. Thus, one would conclude that the intrinsic (unabsorbed) photon
spectrum of this  source in the energy range  shown has an approximate
$E^{-2}$ form.

The  intergalactic $\gamma$-ray  absorption coefficient  (i.e. optical
depth), $\tau(E,z)$, increases monotonically with energy and therefore
leads  to  a steepening  of  the intrinsic  source  spectra  as it  is
observed at  Earth.  For  sources at redshifts  between 0.05  and 0.4,
Stecker \& Scully ~\cite{ss06} have shown that this steepening results
in a well-defined  increase in the spectral index of  a source with an
approximate power-law spectrum in the  0.2 -- 2 TeV energy range. This
increase  is a linear  function in  redshift $z$  of the  form $\Delta
\Gamma = C + Dz$, where  the parameters $C$ and $D$ are constants. The
overall normalization  of the  source spectrum is  also reduced  by an
amount equal  to $ exp \{-(A  + Bz)\}$ ,  again where $A$ and  $B$ are
constants. The values of $A, B, C,$ and $D$ are given for the B and FE
models in Table 1.

These analytic  relations can be to  calculate the intrinsic  0.2 -- 2
TeV power-law  $\gamma$-ray spectra of sources  having known redshifts
in the 0.05  -- 0.4 range and observed  spectral indeces, $\Gamma_{o}$
for both  the B  and FE models  of EBL  evolution Table 2,  taken from
Ref.~\cite{sbs},  gives values  for $\Gamma_{s}$  for  various blazars
obtained using the  formula given in Ref. ~\cite{ss06}.   Table 1 also
shows  the  respective indices  $\Gamma_{e}  =  2\Gamma_{s}-1$ of  the
electron distributions  in the sources  under the assumption  that the
$\gamma$-rays  are produced  by  inverse Compton  interactions in  the
Thomson regime.

Stecker,  Baring  \&  Summerlin~\cite{sbs}  have  also  estimated  the
intrinsic ``isotropic  luminosity.'' of  these blazars, also  shown in
Table  2\footnote{We define  isotropic here  as if  the source  had an
apparent isotropic  luminosity even  though blazars are  highly beamed
and their  flux (and hence their apparent  luminosity) is dramatically
enhanced  by relativistic Doppler  boosting.  This  is similar  to the
nomenclature used for $\gamma$-ray  bursts.  The quantity $\cal{L}$ is
equal  to 4$\pi  \nu F_{\nu}$  given at  h$\nu$ =  1 TeV  in  units of
10$^{36}$ W  (10$^{43}$ erg  s$^{-1}$).}  The isotropic  luminosity of
the blazar sources listed in Table 2 is obtained from the formula
\begin{equation}
  {\cal{L}}     \simeq     4\pi    {{\Gamma_{o}-2}\over{\Gamma_{s}-2}}
(1+z)^{\Gamma_{s}-2} F_{o}[d(z)]^2 e^{(A+Bz)}
\end{equation}
where $d$ is the luminosity distance to the source, and $F_{o}$ is its
observed differential energy  flux at 1 TeV, and  the other factors in
the equation give the  k-correction for the deabsorbed source spectrum
and the normalization correction factor  for absorption, $ exp \{-(A +
Bz)\}$ given in Ref.~\cite{ss06}. The  blazars Mrk 421 and Mrk 501 are
not  included  because their  redshifts  are  significantly less  than
0.05. However, these blazars are analysed in Ref.~\cite{ko03}.

The numbers  given in the last column  of Table 2 are  derived for the
fast evolution  (FE) model. One  may note that  there appears to  be a
trend toward  blazars having flatter intrinsic TeV  spectra and higher
isotropic  luminosities at  higher  redshifts.  However,  one must  be
careful of selection  effects. The TeV photon fluxes  of these sources
as observed  by {\it  H.E.S.S.} and {\it  MAGIC} only cover  a dynamic
range of a  factor of $\sim$20.  Therefore, only  brighter sources can
be observed at higher redshifts. This is because of both diminution of
flux  with  distance  and  intergalactic  absorption~\cite{sds}.   The
observed luminosity-redshift trend is  naturally expected in a limited
population sample spanning a range of redshifts if the TeV-band fluxes
are  pegged  near  an  instrumental  sensitivity  threshold.   A  more
powerful  handle on the  intrinsic spectra  and luminosities  of these
sources  will be  afforded by  the upcoming  {\it  GLAST} $\gamma$-ray
mission, with  its capability for  detecting many blazars  at energies
below 200 GeV.

Table 2 indicates that  some blazars, particularly those observable at
the higher redshifts, appear  to have very hard intrinsic $\gamma$-ray
spectra with indeces in the  range between $\sim$1 and $\sim$1.5.  The
possibility that such spectra  can be obtained from shock acceleration
has not  usually been admitted  when considering properties  of blazar
jets and  their possible emission  spectra~\cite{ah06}.  However, such
spectra can be produced by relativistic shock acceleration~\cite{sbs},
as discussed in  the next section. We will  also discuss observational
evidence  in  the  hard  X-ray  range for  $\gamma$-ray  blazars  with
spectral indeces in the range between $\sim$1 and $\sim$1.5.

\begin{table*}[t]
\begin{center}
{\bf Table  2. Blazar Spectral Indeces  in the 0.2-2  TeV Energy Range
and TeV Luminosities}
\begin{tabular}{cccccc}
&  &  &  &  &  \\   \hline  Source  Name  &  $z$  &  $\Gamma_{obs}$  &
$\Gamma_{source}$(FE $\rightarrow$  B) & $\Gamma_{e}$(FE $\rightarrow$
B) &  $\cal{L}$(1 TeV)  [$10^{36}$ W \\  \hline \hline 1ES  2344+514 &
0.044 & 3.0  & 2.5 $\rightarrow$ 2.6 & 4.0 $\rightarrow$  4.2 & 2.9 \\
Mrk 180 & 0.045 & 3.3  & 2.9 $\rightarrow$ 3.0 & 4.8 $\rightarrow$ 5.0
&  1.2 \\ 1ES1959+650  & 0.047  & 2.7  & 2.3  $\rightarrow$ 2.4  & 3.6
$\rightarrow$  3.8  &  5.4  \\  PKS  2005-489 &  0.071  &  4.0  &  3.4
$\rightarrow$  3.5 & 5.8  $\rightarrow$ 6.0  & 8.6  \\ PKS  2155-304 &
0.117 & 3.3 & 2.2 $\rightarrow$ 2.4 & 3.4 $\rightarrow$ 3.8 & 420 \\ H
2356-309 & 0.165 & 3.1 & 1.5 $\rightarrow$ 1.9 & 2.0 $\rightarrow$ 2.8
& 200  \\ 1ES  1218+30 &  0.182 & 3.0  & 1.2  $\rightarrow$ 1.6  & 1.4
$\rightarrow$  2.2  &  310  \\  1ES  1101-232 &  0.186  &  2.9  &  1.0
$\rightarrow$  1.5 & 1.0  $\rightarrow$ 2.0  & 230  \\ 1ES  0347-121 &
0.188 & 3.1 & 1.2 $\rightarrow$  1.7 & 1.4 $\rightarrow$ 2.4 & 1200 \\
1ES 1101+496 & 0.212 & 4.0 & 1.8 $\rightarrow$ 2.4 & 2.6 $\rightarrow$
3.8 & 930 \\ \hline
\end{tabular}
\end{center}
\end{table*}

\section{Shock Acceleration and Intrinsic Blazar Spectra}

The rapid variability seen in TeV flares drives the prevailing picture
for the blazar source environment,  one of a compact, relativistic jet
that is  structured on small  spatial scales that are  unresolvable by
present $\gamma$-ray telescopes.  Turbulence in the supersonic outflow
in these jets naturally  generates relativistic shocks, and these form
the  principal sites  for acceleration  of electrons  and ions  to the
ultrarelativistic   energies   implied   by   the   TeV   $\gamma$-ray
observations. Within the context of this relativistic, diffusive shock
acceleration mechanism,  numerical simulations  can be used  to derive
expectations for the energy  distributions of particles accelerated in
blazar jets.

Diffusive  acceleration at  relativistic shocks  is less  well studied
than for nonrelativistic flows, yet  it is the most applicable process
for  extreme objects  such as  pulsar winds,  jets in  active galactic
nuclei,   and  $\gamma$-ray   bursts.   A   key   characteristic  that
distinguishes   relativistic   shocks   from  their   non-relativistic
counterparts is  their inherent anisotropy due to  rapid convection of
particles  through and  away  downstream of  the  shock. This  renders
analytic  approaches  more  difficult for  ultrarelativistic  upstream
flows, though advances can be made in special cases, such as the limit
of   extremely  small  angle   scattering  (pitch   angle  diffusion).
Accordingly, complementary  Monte Carlo techniques  have been employed
for   relativistic  shocks   in   a  number   of  studies,   including
test-particle analyses for steady-state shocks of parallel and oblique
magnetic fields  (see refs. in  Ref.  ~\cite{sbs}). This  approach was
used in Ref.~\cite{sbs} to  determine key spectral characteristics for
particles accelerated to high energies at relativistic shocks that are
of  relevance to  acceleration in  blazar  jets. For  a discussion  of
relativistic shock acceleration, see Ref.~\cite{ba04}.

\begin{figure}
  \includegraphics[height=.3\textheight]{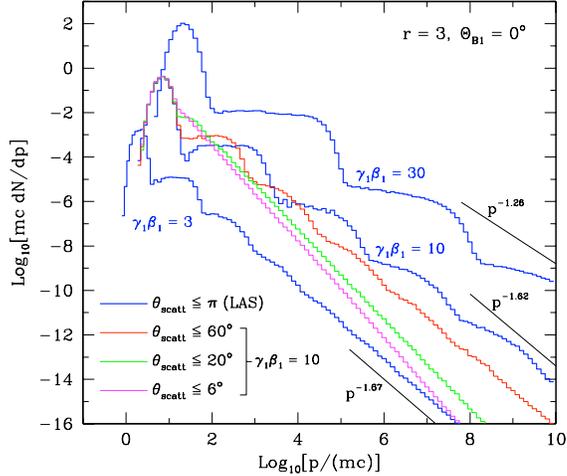}
 \caption{Particle   distribution  functions  $dN/dp$   from  parallel
relativistic shocks obtained from a Monte Carlo simulation of particle
diffusion  and  gyrational  transport.  Scattering  off  hydromagnetic
turbulence is  modeled by randomly  deflecting particle momenta  by an
angle  within  a  cone,  of half-angle  $\theta_{scatt}$,  whose  axis
coincides with  the particle momentum prior to  scattering.  The thick
(blue) lines  are for the  large angle scattering cases  (LAS).  These
asymptotically  approach the  power-laws indicated  by thin  lines, at
high  and  very  high  energies  (not  shown).   Three  smaller  angle
scattering  cases  are  also   shown,  corresponding  to  pitch  angle
diffusion (PAD).  These  have high-energy asymptotic power-law indices
of 1.65, 1.99 and 2.20. }

\label{fig:laspad} 
\end{figure}

Representative  particle momentum distributions  that result  from the
simulation  of  diffusive  acceleration  at  relativistic  shocks  are
depicted  in Figure~\ref{fig:laspad}~\cite{sbs}.   These distributions
are equally applicable to electrons or  ions, and so the mass scale is
not  specified. These  results  highlight several  key features.   The
momentum distributions are flatter when produced by faster shocks with
a larger upstream  flow bulk Lorentz factor of  the jet, $\gamma_{1}$,
given a  fixed velocity compression  ratio.  This is a  consequence of
the  increased  kinematic energy  boosting  occurring at  relativistic
shocks.   Such  a characteristic  is  evident  for  much larger  angle
scattering,  as found  in  the  work of  other  authors referenced  in
Ref.~\cite{sbs}.      What     is     much    more     striking     in
Figure~\ref{fig:laspad} is that the  slope and shape of the nonthermal
particle distribution  depends on the  nature of the  scattering.  The
asymptotic, ultrarelativistic index  of $\Gamma_{e} =2.23$ is realized
only in the mathematical limit of small (pitch) angle diffusion (PAD),
where the particle momentum is stochastically deflected on arbitrarily
small  angular  (and therefore  temporal)  scales.   In practice,  PAD
results  when the  maximum scattering  angle $\theta_{scatt}$  is less
than the Lorentz cone angle 1/$\gamma_{1}$ in the upstream region.  In
such cases, particles diffuse in the region upstream of the shock only
until   their    angle   to    the   shock   normal    exceeds   $\sim
1/\gamma_{1}$. Then they  are rapidly swept to the  downstream side of
the shock. The energy gain per  shock crossing cycle is then roughly a
factor of two, simply derived from relativistic kinematics.

To contrast these  power-law cases, Figure~\ref{fig:laspad} also shows
the  results  given  in  Ref.~\cite{sbs} for  large  angle  scattering
scenarios (LAS,  with $\theta_{scatt} \sim \pi $),  where the spectrum
is highly  structured and much  flatter on average than  $p^{-2}$. The
structure, which becomes  extremely pronounced for large $\gamma_{1}$,
is  kinematic in  origin, where  large angle  deflections lead  to the
distribution   of   fractional   energy   gains  between   unity   and
$\gamma_{1^2}$ in successive shock  transits by particles.  Gains like
this  are kinematically  analogous to  the photon  energy  boosting by
inverse Compton  scattering. Each structured bump  or spectral segment
shown in  Figure~\ref{fig:laspad} corresponds  to an increment  in the
number of shock  crossings, successively from $1\to 3\to  5\to 7$ {\it
etc}. They eventually smooth  out to asympotically approach power-laws
that are indicated  by the lightweight lines in  the Figure.  {\it The
indices of these asymptotic results  are all in the range} $\Gamma_{e}
< 2$. Intermediate cases are also depicted in Figure~\ref{fig:laspad},
with $\theta_{scatt} \sim 4/\gamma_{1}$.  The spectrum is smooth, like
the  PAD case,  but the  index is  lower than  2.23.  Astrophysically,
there is  no reason to exclude  such cases.  From the  plasma point of
view,  magnetic  turbulence  could  easily  be  sufficient  to  effect
scatterings  on this  intermediate  angular scale,  a contention  that
becomes   even  more   salient  for   ultrarelativistic   shocks  with
$\gamma_{1}  \gg 10$.  It  is also  evident that  a range  of spectral
indices  is  produced  when   $\theta_{scatt}$  is  of  the  order  of
1/$\gamma_{1}$. In this case,  the scattering processes corresponds to
a transition between the PAD and LAS limits.

The implications of the numerical simulations given in Ref.~\cite{sbs}
for distributions  of relativistic particles in  blazars are apparent.
There can be a large range in the spectral indices $\Gamma_{e}$ of the
particles  accelerated  in  relativistic  shocks,  and  these  indices
usually differ from $\Gamma_{e} \sim  2.23$. They can be much steeper,
particularly  in  oblique shocks.   However,  they  can  also be  much
flatter,  so  that  quasi-power-law  particle  momentum  distributions
$p^{-\Gamma_{e}}$    with    $\Gamma_{e}    \le   2$    are    readily
achievable. These spectra can be preserved, provided that the electron
spectra produced  during blazar flares are  not significantly affected
by  cooling by synchroton  radiation.  This  requires that  $t_{acc} <
t_{ccol}$ which  constrains both the  magnetic field strength  and the
electron energy  for gyrorsonant acceleration  processes ~\cite{ba02}.
These  electrons  can then  Compton  scatter  to produce  $\gamma$-ray
spectra with $\Gamma < 1.5$.

In fact, recent obsevations of an extreme MeV $\gamma$-ray blazar at a
redshift  of  $\sim$3  by  {\it Swift}~\cite{sa06}  and  the  powerful
$\gamma$-ray quasar PKS 1510-089 at a redshift of 0.361 by {\it Swift}
and {\it Suzaku} ~\cite{ka07}  both exhibited power-law spectra in the
hard X-ray  range which were  significantly harder than  1.5, implying
electron  spectra  significantly  harder   than  value  of  2  usually
considered for shock acceleration. Spectra  in the hard X-ray range do
not  suffer intergalactic  absorption so  that there  is  no ambiguity
concerning their spectral indeces.

\section{Blazar Archaeology}
 
It can be seen from  Figure \ref{tau} that for $\gamma$-ray sources at
the  higher redshifts  there is  a  steeper energy  dependence of  the
optical depth  $\tau (E_{\gamma})$ near  the energy where $\tau  = 1$.
There will  thus be  a sharper absorption  cutoff for sources  at high
redshifts.  It  can easily be seen  that this effect is  caused by the
sharp drop in the ultraviolet photon density at the Lyman limit.

It is expected that {\it GLAST}  will be able to resolve out thousands
of    blazars   from    the    general   extragalactic    $\gamma$-ray
background~\cite{ss96}.   Because of the  strong energy  dependence of
absorption in blazar spectra at  the higher redshifts in the multi-GeV
range, {\it  GLAST} will be able  to probe the  archaeo-IBL and thereby
probe the  early star formation rate.   {\it GLAST} should  be able to
detect blazars at known redshifts $z \sim 2$ at multi-GeV energies and
determine  their  critical  cutoff  energy.   A  simple  observational
technique   for  probing   the   archaeo-IBL  has   been  proposed   in
Ref. \cite{crr}.  In such  ways, {\it GLAST} observations at redshifts
$z \ge 2$ and $E_{\gamma} \sim  10$ GeV may complement the deep galaxy
surveys by probing the {\it total} star formation rate, even that from
galaxies too  faint to be detected  in the deep  surveys.  Future {\it
GLAST} observations in  the 5 to 20 GeV energy range  may also help to
pin down  the amount of  dust extinction in high-redshift  galaxies by
determining  the mean  density of  ultraviolet photons  at  the higher
redshifts through their absorption  effect on the $\gamma$-ray spectra
of  high redshift  sources.   If the  diffuse $\gamma$-ray  background
radiation is  from unresolved blazars \cite{ss96},  a hypothesis which
can  be independently  tested by  {\it GLAST}  \cite{ss99}, absorption
will steepen  the spectrum of this radiation  at $\gamma$-ray energies
above $\sim  10$ GeV  \cite{ss98}. Thus, {\it  GLAST} can  also aquire
information about the evolution of the IBL in this way.

Conversely,  observations   of  sharp  high  energy   cutoffs  in  the
$\gamma$-ray  spectra of  sources  at unknown  redshifts  can be  used
instead of spectral lines to give a measure of their redshifts.

Table 2 indicates  that there may be an evolution  of TeV blazars with
redshift, both in  luminosity and spectral index. It  may be that past
blazars had jets with higher bulk Lorentz factors than low-redshift BL
Lacs. It may also be that these apparent trends are only the result of
selection effects. There is only  a small sample of ten sources listed
in Table 2.  {\it GLAST} will  be able to explore this question, as it
has the potential of seeing thousands of blazars~\cite{ss96}.

\section{Extragalactic Astronomy: The Greatest Dig in the Universe}

In keeping with the cosmic high energy theme of the 30th
International Cosmic Ray Conference, the emphasis in this paper has
been on high energy photons ($\gamma$-rays). However, we have seen
the intimate connection between these photons and the low energy
photons that are the subjects of more traditional astronomy. In fact,
radio surveys, ongoing infrared-to-ultraviolet deep astronomical surveys,
and surveys of the more newly discovered extragalactic $\gamma$-ray
sources and $\gamma$-ray bursts, are most effectively used when considered
together to create a consistent and compelling picture of the past
history of the Universe. 

\section{Epilogue}

In memory of Frank Culver Jones, my onetime mentor and
friend for four decades. Frank passed away earlier this year.
He only missed one ICRC conference since 1961. He also served as
secretary to the ICRC. He will be missed by the cosmic-ray
community.


\end{document}